# Topological insulator Bi$_2$Se$_3$ films on rare earth iron garnets and their high-quality interfaces

C. C. Chen (陳俊嘉),[1] K. H. M. Chen (陳可璇),[1] Y. T. Fanchiang (范姜宇廷),[2] C. C. Tseng (曾俊智),[1] S. R. Yang (楊尚融),[1] C. N. Wu (巫啓男),[1] M. X. Guo (郭孟鑫),[1] C. K. Cheng (鄭兆凱),[2] C. T. Wu (吳建霆),[3] M. Hong (洪銘輝),[2, a] and J. Kwo (郭瑞年)[1, a]

[1] Department of Physics, National Tsing Hua University, Hsinchu 30013, Taiwan

[2] Department of Physics, National Taiwan University, Taipei 10617, Taiwan

[3] National Nano Device Laboratories, Hsinchu 30013, Taiwan

a) To whom the correspondence is addressed: raynien@phys.nthu.edu.tw (J. Kwo) and mhong@phys.ntu.edu.tw (M. Hong)

## Abstract

The integration of quantum materials like topological insulators (TIs) with magnetic insulators (MIs) has important technological implications for spintronics and quantum computing. Here we report excellent crystallinity of c-axis oriented epitaxial TI films Bi$_2$Se$_3$ grown on MI films, a rare earth iron garnet (ReIG), such as thulium iron garnet (Tm$_3$Fe$_5$O$_{12}$, TmIG) by molecular beam epitaxy (MBE) with a Se-buffered low-temperature (SBLT) growth technique. We demonstrated a streaky reflection high-energy electron diffraction pattern starting from the very first quintuple layer of Bi$_2$Se$_3$, indicating the high-quality interface between TmIG and Bi$_2$Se$_3$, a prerequisite for studying interfacial exchange coupling effects. The strong interfacial exchange interaction was manifested by observations of anomalous Hall effect in the Bi$_2$Se$_3$/TmIG bilayer and a shift of ferromagnetic resonance field of TmIG induced by Bi$_2$Se$_3$. We have reproducibly grown high-quality Bi$_2$Se$_3$/ReIG and interfaces using this new TI growth method, which may be applied to grow other types of van der Waals (vdW) hetero-structures.

Topological insulators (TIs) have attracted great interests in the field of condensed matter sciences because of the exotic topological edge or surface state (TSS) arising from the bulk band inversion driven by strong spin-orbit coupling (SOC).[1] The electrons at the surface state are protected by time reversal symmetry (TRS), and the spin-momentum locking feature has led to a transport forbidding back-scattering.[2] A series of three-dimensional (3D) TIs with a single Dirac cone at Γ point of momentum spaces, such as $Bi_2Se_3$, $Bi_2Te_3$, and $Sb_2Te_3$, were discovered and received feverish studies in the past decade.[3,4] Recently, research interests have turned to the phenomena emerging from broken TRS in ferromagnetic TIs. Quantum anomalous Hall effect (QAHE), the latest member of the Hall family, exhibits a quantized Hall conductivity $\sigma = \frac{e^2}{h}$ without any external magnetic field, and generates exciting new physics to be explored in the ferromagnetic TI systems.[5] The dissipationless chiral transport in QAH states displays high potential in the development of spin transport devices with low-power consumption.[6,7] Although QAHE was demonstrated in magnetically doped TI systems, the QAH regime is restricted to be below 2 K.[8,9] In addition, extra carriers and inevitable crystal defects were generated by excess dopants in the films, which may destroy novel quantum states.[10]

An alternative way to introduce ferromagnetism in the TI is to couple the TI with a magnetic insulator (MI), thereby inducing a magnetic order through magnetic proximity effect (MPE). Compared to the samples with magnetic doping, a more uniform magnetization without crystal defects may be developed in the hetero-structures under MPE. The ferromagnetic insulators are preferred over ferromagnetic metals to ensure a clean interface, avoiding chemical interaction and current shunting. A TI/MI hetero-structure with a small lattice mismatch was first demonstrated in $Bi_2Se_3$/EuS.[11] The epitaxial growth has resulted in the persistence of MPE to room temperature, suggesting that much stronger exchange field can be generated to break TRS.[12] However, the low Curie temperature ($T_c$) ~17 K and the preferred in-plane shape anisotropy of EuS thin films precluded a large out-of-plane magnetization in absence of magnetic fields for practical applications.

In contrast, ferrimagnetic rare earth iron garnets (ReIGs) of high $T_c$ above 500 K may be an attractive choice. Thulium iron garnet ($Tm_3Fe_5O_{12}$ (TmIG)) exhibiting perpendicular magnetic anisotropy (PMA) is preferred over the well-known yttrium iron garnet ($Y_3Fe_5O_{12}$ (YIG)) with in-plane anisotropy (IMA). To grow high-quality TI thin films on ReIG substrates, however, is challenging, since the large lattice mismatch and the complicated atomic arrangement on the garnet surface have posed difficulties in developing a sharp interface, the prerequisite to establish a direct coupling effect between TSS and MI. In most cases, an interlayer with poor crystallinity formed in TI/ReIG bilayers,[13-16] which would greatly suppress the interfacial coupling. This, however, is inevitable if the TI films were grown by using the conventional two-step method.[17] Although an MPE above 400 K has been recently demonstrated in $(Bi,Sb)_2Te_3$/TmIG,[18] the TI/ReIG interface quality showed notable variations, leading to a large fluctuation of MPE properties and hampering further systematic studies.[16,19] Therefore, it is of upmost importance to establish a robust experimental procedure to consistently obtain high-quality TI/ReIG samples.

Here, we have developed a new MBE growth method to overcome the large lattice mismatch in the TI/MI hetero-structures, enabling excellent crystallinities at the hetero-interface and the bulk of TI. Our method has *reproducibly* produced high-quality $Bi_2Se_3$/TmIG hetero-structures, with an atomically sharp interface. The MBE-grown $Bi_2Se_3$ thin films were deposited on tensile-strained TmIG films of PMA [20,21] with a high $T_c$ ~560 K, which is expected to fulfill the low-power-dissipation magnetization switching.[22] The $Bi_2Se_3$/TmIG thin films with a sharp interface showed excellent crystallinity characterized by several structural analyses. Hysteresis loops were observed in low temperature magnetoresistance (MR), which may correspond to anomalous Hall effect (AHE) coming from the magnetized interface on the TI side, or efficient spin transfer at the interface analogous to the spin Hall magnetoresistance (SMR) observed in heavy metal/MI.[23] Moreover, a large shift of the ferromagnetic resonance (FMR) field was observed, indicating a change of magnetic anisotropy induced by TI. The structural perfectness has contributed to clear observations of the strong interfacial exchange coupling at the TI/MI interface. This has enabled us to further investigate MPE and spin-transfer mechanism in TI/MI structures, where MI may include TmIG, YIG, and $Fe_3O_4$[24], etc. Furthermore, similar growth can be applied to other 2D materials with van der Waals (vdW) bonding to substrates.

The TmIG films were grown on gadolinium gallium garnet ($Gd_3Ga_5O_{12}$ (GGG)) substrates by off-axis sputtering.[20,21] The TmIG films were *ex-situ* transferred into the MBE system immediately after they were loaded out from the sputtering chamber. Our MBE system has a base pressure around $4 \times 10^{-10}$ Torr with high-purity (99.9999%) Bi and Se sources in standard effusion cells. Bright and clear reflection high-energy electron diffraction (RHEED) patterns of TmIG (111) after outgassing the sample at 150 °C indicated the attainment of a high degree of crystallinity and smooth surface, ensuring the growth of high-quality hetero-structures (Fig. 1(a)). The growth rate of our $Bi_2Se_3$ films is 0.3 - 0.4 nm/min and the ratio of Se/Bi flux for the growth is 5. Before the growth of TI films, the garnet surface was first covered by an amorphous Se buffer layer ~2 nm thick, followed by an amorphous $Bi_xSe_{1-x}$ layers ~1 nm thick at 50 ℃. As the substrate temperature ascended to 250 ℃, the amorphous mixture of $Bi_xSe_{1-x}$ crystallized and exhibited streaky $Bi_2Se_3$ RHEED patterns, revealing an excellent crystallinity of the first quintuple layer (QL) forming at the interface (Fig. 1(b)).

In sharp contrast, using the conventional two-step growth procedure, spotted ring RHEED features of the first QL indicated polycrystalline $Bi_2Se_3$ domains with a rough surface forming at the interface (Fig. 1(e)). The complicated atomic arrangement of garnet surface prevented Bi and Se atoms from stacking orderly on such surface. The distinct difference revealed at the first QL between the two different growth methods showed that the crystallinity of the initial layer was significantly improved by our new growth technique, which is crucial for fabricating TI/MI hetero-structures.

The Se buffer layer is key to the initiation of the crystallization process of $Bi_2Se_3$, which would otherwise be hampered by the lattice-mismatched substrate hosting dangling bonds. The Se buffer layer was covered merely by about 1 nm thick $Bi_2Se_3$. It is conceivable that the excess Se of the buffer layer evaporated away as the substrate temperature was raised to 250 ℃, leaving a well crystalized ~1 nm $Bi_2Se_3$ which became an ideal template for the rest of film growth (Fig. 2(c)). The absence of excess Se was further confirmed by our high-resolution transmission electron microscope (HRTEM) image in Fig. 3. The effect of the Se buffer layer thickness on the quality of the first QL of $Bi_2Se_3$ was also investigated, and the results are given in the Supplemental Material.

More distinct RHEED patterns of $Bi_2Se_3$ films in further growth by our recipe have shown that the crystallinity of $Bi_2Se_3$ were steadily enhanced during the entire growth (Fig. 1(c), (d)). Very bright RHEED patterns were obtained toward the end of the growth. On the other hand, for the two-step growth, the 2D line features did not show up among the rings until the 3$^{rd}$ QL (Fig. 1(f)). The line features finally took over, leading to streaky RHEED patterns at the 7$^{th}$ QL (Fig. 1(g)). From the evolution of RHEED patterns, the grains began to enlarge at the 3$^{rd}$ QL and started to form a smoother 2D surface. In contrast to the two-step growth, the streaky RHEED of the 1$^{st}$ QL using our growth technique is evidenced for a well ordered atomic arrangement at the interface, which is required to form a strong coupling of the $Bi_2Se_3$ bottom surface with the MI underneath. Hereafter, we refer our new growth procedure as "Se-buffered low-temperature" (SBLT) growth. In addition, two different diffraction spacing, $[11\bar{2}0]$ and $[1\bar{1}00]$, existed together in the RHEED patterns of $Bi_2Se_3$, which have also been observed in previous works of TIs grown on garnet structures.[13-15,18,19,25] The intensities of streaks were almost the same in all in-plane angles, corresponding to domains randomly oriented azimuthally on the surface. Moreover, the X-ray diffraction (XRD) results on the SBLT-grown $Bi_2Se_3$ films (in the Supplemental Material) showed no orientation relationship with regard to the substrate, implying that the $Bi_2Se_3$ domains at the interface crystallized by themselves instead of epitaxially grew on the substrate. The overall SBLT growth process is illustrated graphically in Fig. 2(a)-(d), and the distinct difference in surface morphologies between the films grown by the SBLT and the conventional two-step methods is given in the Supplemental material.

We further characterized the interfacial atomic stacking of $Bi_2Se_3$/TmIG using HRTEM. Figure 3 shows the layered structure of $Bi_2Se_3$, where each QL is comprised of Se-Bi-Se-Bi-Se atomic sequence.

As visualized from the interface, the stacking feature emerged from the first QL is without an extra layer, providing direct evidence that virtually no extra Se atoms, if any, remained after the sample was annealed at 250 °C. Further, we have employed high-angle annular dark-field (HAADF) imaging using Cs-corrected STEM to give the interface atomic arrangement with a higher resolution. As shown in the inset of Fig. 3, a darker interfacial layer, presumably the first QL of $Bi_2Se_3$, was observed, which exhibits a similar crystal structure as the well-crystalized part of the $Bi_2Se_3$ film. The $Bi_2Se_3$-like structure may evolve from the $Bi_xSe_{1-x}$ seed layer and containing smaller crystal domains compared to those in the following QLs. In addition, at the initial growth stage a Se buffer was used to suppress direct chemical bonds between $Bi_2Se_3$ and TmIG to promote vdW epitaxy. The re-evaporation of the Se buffer layer and the annealing to 250 °C potentially modified the chemical environment, which may enhance the interfacial bonding and diminish the vdW nature of the first QL. In such a scenario, the first QL acts as a transition layer to connect the 2D ($Bi_2Se_3$) and 3D (TmIG) structures.[26] True vdW epitaxy can be realized when the substrate is of layered structure such as graphene, monolayer $MoS_2$ or $In_2Se_3$.[27-29] However, our study suggests that, for a non-vdW type substrate, a transition layer may be necessary for obtaining a high-quality TI-based hetero-structure.

Electrical transport measurements on the $Bi_2Se_3$/TmIG bilayer were conducted. In Fig. 4(a), we observed the feature of AHE: a clear hysteresis loop in the Hall resistance ($R_{xy}$) after subtracting the ordinary Hall effect background of the $Bi_2Se_3$/TmIG. The coercive field $H_c$ of the hysteresis loop increased as the temperature decreased, while the AHE amplitude $R_{AHE}$ was weakly temperature dependent from 50 K to 220 K. Above 220 K, a hysteresis loop signal was too small to detect due to an insufficient signal-to-noise ratio. The increasing $H_c$ toward low temperature might be associated with the strengthened PMA of TmIG films because of the increasing film strains at low temperatures.[30] There are two possible origins of the AHE signals. Firstly, if the $Bi_2Se_3$ bottom surface was magnetized by MPE, the induced magnetization would contribute to the $R_{xy}$. We note that the temperature dependence of $R_{AHE}$ for our $Bi_2Se_3$/TmIG is distinct from the earlier work,[25] where the $R_{AHE}$ of $(Bi_{0.16}Sb_{0.84})_2Te_3$/YIG decreased with increasing temperature and vanished at 150 K. Moreover, robust AHE loops persisting up to 400 K were reported in $(Bi,Sb)_2Te_3$/TmIG.[18] The different $T$ dependence of $R_{AHE}$ of our $Bi_2Se_3$/TmIG films was attributed to the bulk states of $Bi_2Se_3$ being conductive, where the transport properties of the magnetized TI may entangle with the bulk carriers and hence the contribution from the interface was suppressed. The second possible mechanism of the AHE is based on the spin current transport in $Bi_2Se_3$. An applied charge current in $Bi_2Se_3$ can generate spin accumulation at the interface by spin Hall effect in the bulk or Rashba-Edelstein effect of the surface states. Such non-equilibrium spins would modulate the conductance of TI films if efficient spin-transfer occurs at the interface.[31] Nonetheless, both MPE and spin transfer at the interface originate from the interfacial exchange coupling, and the observation of AHE attests to the high quality of our $Bi_2Se_3$/TmIG sample.

As an independent probe of the interfacial effect, FMR measurements were performed to look into the magnetization dynamics of TmIG under the influence of $Bi_2Se_3$. Figure 4(b) and (c) compare the resonance field ($H_{res}$) of bare TmIG and $Bi_2Se_3$(7 nm)/TmIG prepared by the two-step method and the SBLT growth method under an in-plane external field. A clear negative shift of $H_{res}$ is observed after the growth of $Bi_2Se_3$ layer, and in particular, the $H_{res}$ induced by the $Bi_2Se_3$ film grown by the SBLT method in Fig. 4(c) is almost three times larger than that presented in Fig. 4(b). The negative shift of $H_{res}$ corresponds to an enhanced IMA at the interface as reported in our previous work on $Bi_2Se_3$/YIG.[15] Based on the study, the magnetization dynamics of TmIG was likely profoundly affected by the TSS of $Bi_2Se_3$. The magnetization dynamics modulated by $Bi_2Se_3$ can thus be viewed as a consequence of strong interfacial exchange coupling as well as spin-momentum locking of TSS.[16] Note that, despite that $Bi_2Se_3$/TmIG exhibits strain-induced PMA from the bulk of TmIG, $Bi_2Se_3$ still induces a large interfacial IMA, representing an intensive interfacial exchange interaction existing in the $Bi_2Se_3$/TmIG bilayer prepared by the SBLT method.

In conclusion, we have invented a new SBLT growth method of attaining high-quality $Bi_2Se_3$ thin films on ReIG by adding an ultrathin Se buffer layer prior to the $Bi_2Se_3$ deposition. Corroborated by extensive structural analyses, the Se buffer effectively isolated the crystallization process of $Bi_2Se_3$ from the rather complicated atomic structure of the ReIG surface, and initiated the layered structure of $Bi_2Se_3$ film at the very first QL. The excellent crystallinity of the interfacial layer is essential to give high-quality growth for thicker $Bi_2Se_3$ films. This is testified by the observation of clear AHE in $Bi_2Se_3$/TmIG prepared by the SBLT growth. Although the physical origin of the AHE by MPE or spin current transport remains to be clarified, both mechanisms pointed to the existence of an enhanced interfacial exchange coupling. An independent evaluation of the enhanced exchange coupling by means of FMR has observed notably larger in-plane magnetic anisotropy in better-quality of $Bi_2Se_3$, manifested as a shift of ferromagnetic resonance field compared to that of bare TmIG. Lastly, it is noteworthy that the new method of growing $Bi_2Se_3$ can be further extended to other TIs of similar crystal-chemical properties such as $Bi_2Te_3$ and bulk insulating $(Bi,Sb)_2Te_3$, and even other type of layered material suitable for vdW epitaxy.

**Acknowledgement**

The authors would like to thank J. S. Wei for the technical support in transport measurements. Technical support from the Nano group Public Laboratory, Institute of Physics, Academia Sinica in Taiwan is acknowledged. This work was supported by the Ministry of Science and Technology, Taiwan, with grant numbers MOST 105-2112-M-007-014-MY3, 106-2112-M-002-010-, and 106-2622-8-002-001.

**Figure captions**

Fig. 1 RHEED patterns of (a) TmIG(111) surface; after growing (b) 1 QL; (c) 3 QL; (d) 7 QL of $Bi_2Se_3$ with a thin Se buffer layer; after growing (e) 1 QL; (f) 3 QL; (g) 7 QL of $Bi_2Se_3$ by two-step method.

Fig. 2 Illustrations for the SBLT growth. (a) First covering a Se buffer layer ~2 nm thick on TmIG; (b) Depositing ~1 nm amorphous mixture of Bi and Se; (c) $Bi_2Se_3$ seed layer crystallizing and excess Se evaporating during the substrate temperature increasing; (d) Further growth of $Bi_2Se_3$ films on the seed layer.

Fig. 3 HRTEM image of 12 nm $Bi_2Se_3$ on TmIG. Inset: Cs-corrected HAADF-STEM image showing the interface fine structure.

Fig. 4 (a) AHE signals were observed in 7 nm $Bi_2Se_3$/TmIG structure from 70 K to 220 K; (b) FMR signals before (black) and after (blue and red) deposition of $Bi_2Se_3$ on TmIG with the SBLT growth.

Fig. 1

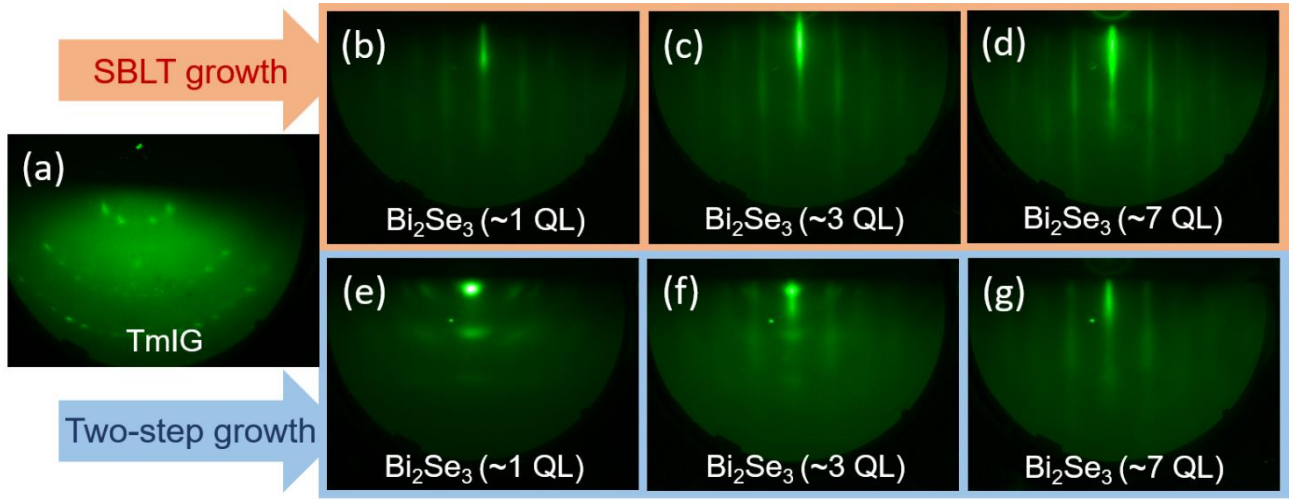

Fig. 2

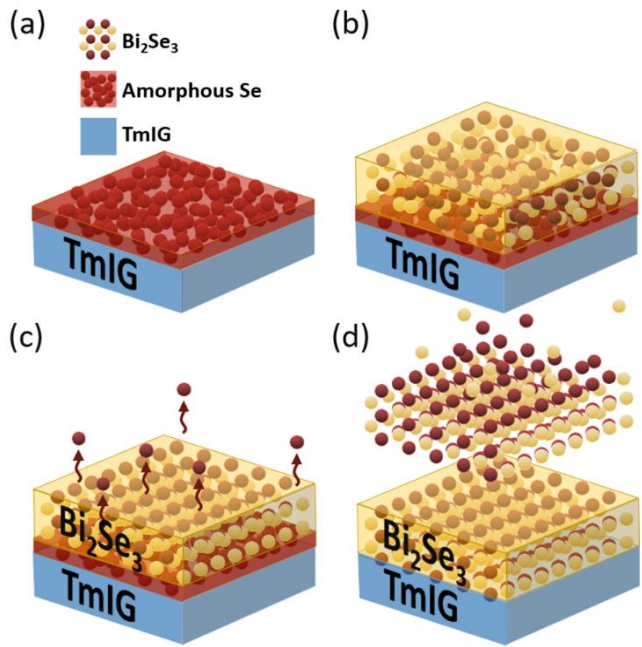

Fig. 3

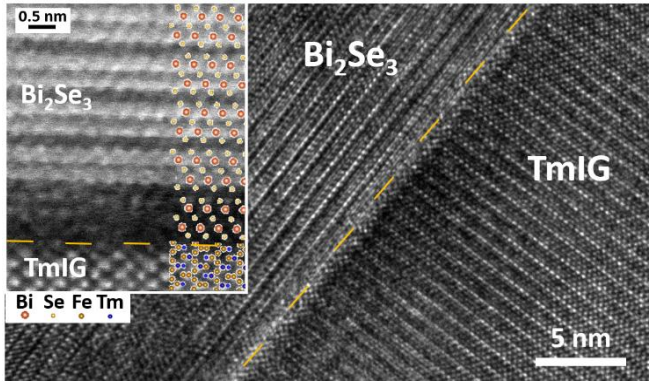

Fig. 4

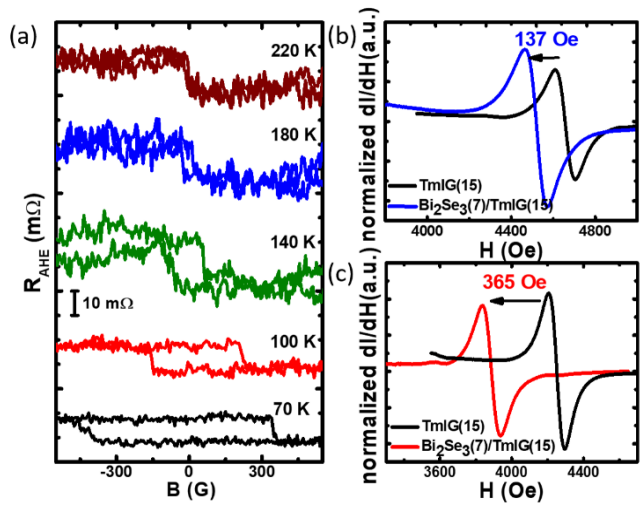